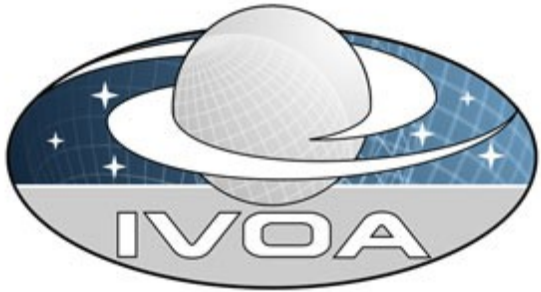

*I*nternational

*V*irtual

*O*bservatory

*A*lliance

# IVOA Data Access Layer Interface Version 1.0

## IVOA Recommendation 2013-11-29

**Interest/Working Group:**

**Editors:**

**Patrick Dowler**

**Authors:**

**Patrick Dowler, Markus Demleitner, Mark Taylor, Doug Tody**

# Abstract

This document describes the Data Access Layer Interface (DALI). DALI defines the base web service interface common to all Data Access Layer (DAL) services. This standard defines the behaviour of common resources, the meaning and use of common parameters, success and error responses, and DAL service registration. The goal of this specification is to define the common elements that are shared across DAL services in order to foster consistency across concrete DAL service specifications and to enable standard re-usable client and service implementations and libraries to be written and widely adopted.

# Status of This Document

This document has been produced by the Data Access Layer Working Group.

*It has been reviewed by IVOA Members and other interested parties, and has been endorsed by the IVOA Executive Committee as an IVOA Recommendation. It is a stable document and may be used as reference material or cited as a normative reference from another document. IVOA's role in making the Recommendation is to draw attention to the specification and to promote its widespread deployment. This enhances the functionality and interoperability inside the Astronomical Community.*

*A list of current IVOA Recommendations and other technical documents can be found at http://www.ivoa.net/Documents/.*

# Acknowledgements

The authors would like to thank all the participants in DAL-WG discussions for their ideas, critical reviews, and contributions to this document.

# Contents

# 1 Introduction

The Data Access Layer Interface (DALI) defines resources, parameters, and responses common to all DAL services so that concrete DAL service specifications need not repeat these common elements.

## 1.1 The Role in the IVOA Architecture

DALI defines how DAL service specifications use other IVOA standards as well as standard internet designs and protocols.

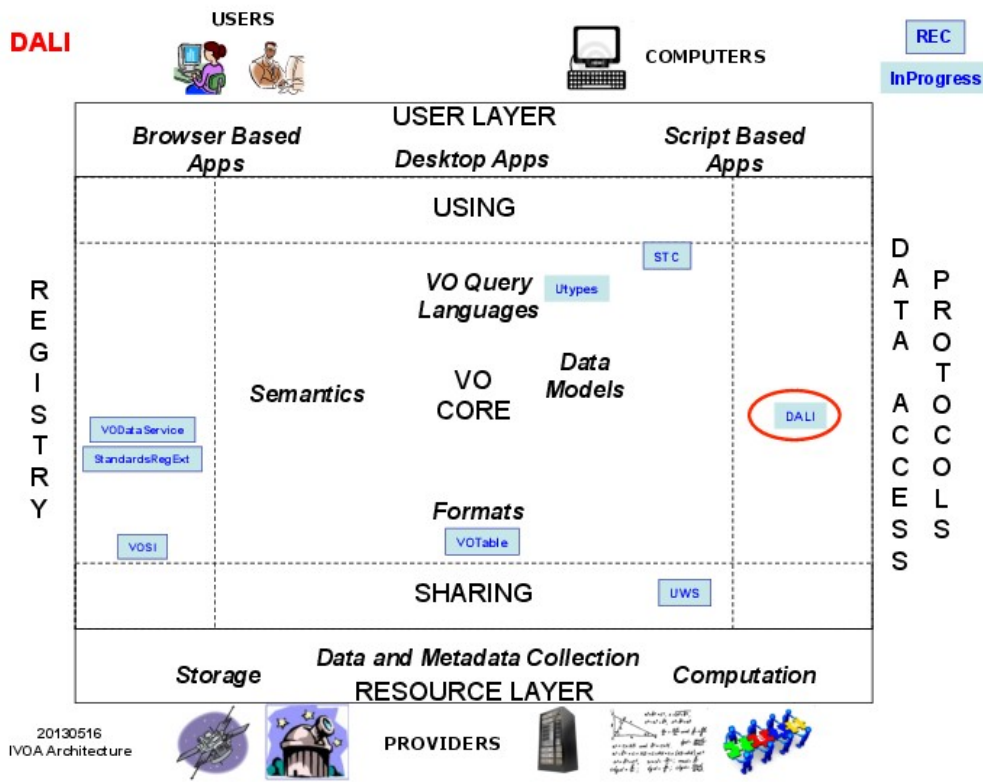

DAL services use the Universal Worker Service (UWS) pattern for asynchronous requests. All DAL services include Virtual Observatory Support Interfaces (VOSI) resources. DAL services generally use VOTable as the default output format for successful tabular output and to return error documents.

Astronomical coordinate values accepted and returned by DAL services use a string representation of the Space-Time Coordinates (STC) data model. The concrete DAL service specification defines whether the returned resources are serializations of a particular standard data model. For preserving backwards compatibility or to enable service-specific use cases, the concrete DAL service specification may explicitly specify the use of ad-hoc Utypes.

A registry extension schema, usually extending VODataService, is needed to describe the capabilities of a DAL service. This schema is used within the VOSI-capabilities resource and in registry records for the service.

## 1.2  Example Usage of DALI Specification

The DALI specification defines common elements that make up Data Access Layer (DAL) services. DAL service specifications will refer to the sections in this document by name rather than include all the explanatory text. For example, suppose a document defines a service that stacks FITS images asynchronously. The specification could say that the service has the following resources:

- a DALI-async resource that accepts one or more UPLOAD parameters (Section  3.2.5 ) where the resources are FITS images; the resource could also define a fixed set of error messages for anticipated failure modes

- a VOSI-availability resource (Section  2.4 )

- a VOSI-capabilities resource (Section   2.5 ) conforming to a specified registry extension schema

and would have to define the registry extension schema to be used to register services and to implement the VOSI-capabilities resource. Most of the service specification would be in defining the semantics (possibly controllable via additional input parameters) of the computations to be performed and in defining the extension schema to describe service functionality and limits (e.g. maximum input or result image sizes, result retention time and policies). The registry extension schema may be part of the service specification or a separate document.

## 2 Resources

DAL services are normally implemented as HTTP REST [16] web services, although other transport protocols could be used in the future[1]. The primary resource in a DAL service is a job. A DAL job is defined by parameters (Section  3 ) and can be executed either synchronously or asynchronously. A concrete service specification defines the job parameters and the manner of execution is defined by separate resources below.

In addition to job list resources, DAL services also implement several Virtual Observatory Support Interface (VOSI) resources to describe service availability, capabilities, and content.

A concrete DAL service **must** define at least one DALI-async or DALI-sync resource. It may define both with the same job semantics (e.g. TAP-1.0 [13]) or it may define one with one kind of job and the other with a separate kind of job (a service that does some things synchronously and others asynchronously).

The following table summarises the resources that are required in all concrete DAL service specifications (and thus in all DAL services) and which kinds of resources are defined and specified as required or optional in a concrete specification.

| resource type | resource name | required |
|---|---|---|
| DALI-async | service specific | service specific |
| DALI-sync | service-specific | service specific |
| DALI-examples | /examples | no |
| VOSI-availability | /availability | yes |
| VOSI-capabilities | /capabilities | yes |
| VOSI-tables | /tables | service specific |

A simple query-only DAL service like ConeSearch can be easily described as having a single DALI-sync resource where the job is a query and the response is the result of the query.

### 2.1 Asynchronous Execution: DALI-async

Asynchronous resources are resources that represent a list of asynchronous jobs as defined by the Universal Worker Service (UWS) pattern [10]. Requests can create, modify, and delete jobs in the job list. UWS also specifies special requests to modify the phase of the job (cause the job to execute or abort).

---

1 We do describe only RESTful HTTP web services here, but a future version of this specification may make HTTP optional and allow for other transport protocols.

As specified in UWS, a job is created by using the HTTP POST method to modify the job list. The response will always be an HTTP redirect (status code 303) and the Location (HTTP header) will contain the URL to the job (a child resource of the job list).

```
POST http://example.com/base/async-jobs
```

The response will include the HTTP status code 303 (See Other) and a header named Location with a URL to the created job as a value, for example:

```
Location: http://example.com/base/async-jobs/123
```

The job description (an XML document defined by the UWS schema) can always be retrieved by accessing the job URL with the HTTP GET method:

```
GET http://example.com/base/async-jobs/123

<?xml version="1.0" encoding="UTF-8"?>
<uws:job xmlns:uws="http://www.ivoa.net/xml/UWS/v1.0">
  <uws:jobId>123</uws:jobId>
  <uws:runId>test</uws:runId>
  <uws:ownerId xsi:nil="true" />
  <uws:phase>PENDING</uws:phase>
  <uws:quote>2013-01-01T12:34:56</uws:quote>
  <uws:startTime/>
  <uws:endTime/>
  <uws:executionDuration>600</uws:executionDuration>
  <uws:destruction>2013-02-01T00:00:00</uws:destruction>
  <uws:parameters>
    <uws:parameter id="LANG">ADQL</uws:parameter>
    <uws:parameter id="REQUEST">doQuery</uws:parameter>
    <uws:parameter id="QUERY">select * from tab</uws:parameter>
  </uws:parameters>
  <uws:results/>
</uws:job>
```

In addition to the UWS job metadata, DAL jobs are defined by a set of parameter-value pairs. The client may include parameters in the initial POST that creates a job or it may add additional paramaters by a POST to thecurrent list of parameters, for example:

```
http://example.com/base/async-jobs/123/parameters
```

DALI-async resources may provide other ways to interact with jobs as specified in current or future UWS specifications, with the following exception: the UWS-1.0 standard may be interpreted to allow POSTing of job parameters to the job URL, but DALI-async resources **must** not accept job parameters at this URL.

Job parameters may only be POSTed while the job is in the PENDING phase; once execution has been requested and the job is in any other phase, job parameters may not be modified.

A concrete DAL service specification will specify zero or more asynchronous job submission resources and whether they are mandatory or optional. It may mandate a specific resource name to support simple client use, or it can allow the resource name to be described in the service metadata (Section 2.5 ).

## 2.2 Synchronous Execution: DALI-sync

Synchronous resources are resources that accept a request (a DAL job description) and return the response (the result) directly. Synchronous requests can be made using either the HTTP GET or POST method. If a specific type of job is exposed through both DALI-async and DALI-sync resources (e.g. TAP queries), then the parameters used to specify the job are the same for this pair of (synchronous and asynchronous) jobs. Service specifications may also specify different types of jobs on different resources, which would have different job parameters.

A synchronous job is created by a GET or POST request to a synchronous job list, executed automatically, and the result returned in the response. The web service is permitted to split the operation of a synchronous request into multiple HTTP requests as long as it is transparent to standard clients. This means that the service may use HTTP redirects (status code 302 or 303) and the Location header to execute a synchronous job in multiple steps. For example, a service may

- immediately execute and return the result in the response, or
- the response is an HTTP redirect (status code 303) and the Location (HTTP header) will contain a URL; the client accesses this URL with the HTTP GET method to execute the job and get the result

Clients must be prepared to get redirects and follow them (using normal HTTP semantics) in order to complete requests.

A concrete DAL service specification will specify zero or more synchronous job submission resources and whether they are mandatory or optional. It may mandate a specific resource name to support simple client use, or it can allow the resource name to be described in the service capability metadata (Section 2.5 ).

## 2.3 Examples: DALI-examples

The DALI-examples resource returns a document with usage examples or similar material to the user. In DAL services, this resource is always accessed as a resource named *examples* that is a child of the base URL for the service. The following specification is intended to make sure the content is usable for both machines and humans. As such, the DALI-examples resource contains additional markup conforming to the RDFa 1.1 Lite [17] specification, which defines the following attributes: vocab, typeof, property, resource, and prefix (although we do not include any use of the prefix attribute).

The DALI-examples capability identifier is:

```
ivo://ivoa.net/std/DALI#examples
```

DAL services may implement the */examples* resource and include it in the capabilities described by the VOSI-capabilities resource (Section 2.5 ); if they do not, retrieving its URL **must** yield a 404 HTTP error code.

The document at */examples* **must** be well-formed XML. This restriction is imposed in order to let clients parse the document using XML parsers rather than the much more complex parsers (e.g. HTML5 parsers). It is therefore advisable to author it in XHTML, although this specification does not prescribe any document types.

The document should be viewable with "common web browsers". Javascript or CSS must not be necessary to find and interpret the elements specified below. Apart from that, service operators are free to include whatever material or styling they desire in addition and within the example elements defined here.

The elements containing examples **must** be descendants of an element that has a *vocab* attribute with the value equal to the DALI-examples capability identifier (above),  for example:

```
<div vocab="ivo://ivoa.net/std/DALI#examples">
...
</div>
```

No other *vocab* attributes are allowed in the document. Each example resides in an element that has a *typeof* attribute with the value e*xample*. All such elements **must** have an *id* attribute to allow external referencing via fragments and a *resource* attribute with a reference pointing to the element itself. As an example,

```
<div id="x" resource="#x" typeof="example"> ... </div>
```

located inside the vocab element (the one with the *vocab* attribute) would contain an example referrable via the x fragment identifier. The *div* element is a suitable HTML element to hold an *example*.

The content of the example is expressed using the *property* attribute. For DALI-examples, we define the following values for the *property* attribute: *name*, *capability*, *generic-parameter*, and *continuation*. Each example **must** include one *name.* An example should contain one *capability* and **must** include it if the service supports multiple capabilities and examples would be ambiguous. Examples for services that support a single capability (e.g. Simple Cone Search) do not need to include the capability property, although doing so makes the examples more easily machine-readable. The value of the property is the content of the element. For the *generic-parameter*, we also define a value for the *typeof* attribute (see below) as the content of the element has a defined structure; the generic-parameter property is used to describe parameters for invoking RESTful web service capabilities. DAL service specifications may define additional properties so they can mark up additional information in their examples.

In principle, any element permitted by the document type can include the RDFa attributes, so authors may re-use existing markup intended for display. Alternatively, the span element is a good choice when the example values are included in surrounding text and the author does not want any special rendering to be applied by the machine-readable additions.

### 2.3.1  name property

The content of this element should be suitable for display within a space-limited label in user interface and still give some idea about the meaning of the example. In XHTML, a head element (*h2*, say) would usually be a good choice for the example name, for example:

```
<h2 property="name">Synchronous TAP query</h2>
```

### 2.3.2  capability property

The capability property for an example specifies which service capability the example is to be used with. For example, if the text is describing how to use an SSA-1.1 service, the example could contain:

```
<span property="capability">ivo://ivoa.net/std/SSA/v1.1</span>
```

IVOA standard service capabilities are defined as URIs, so example documents may want to show the URI or show more user-friendly text depending on the expected audience for the document. For specifications that do not define specific capability identifiers, the IVORN for the specification itself should be used.

### 2.3.3  generic-parameter property

Request parameters are included within the example by using the *generic-parameter* property. The element **must** also be assigned a *typeof* attribute with value of *keyval*. Within this element, the document **must** include a pair of elements with a property named key and value, where the plain-text content are the parameter name and value respectively. Multiple generic-parameter(s) are permitted, for example:

```
<span property="generic-parameter" typeof="keyval">

   <span property="key">REQUEST</span>

   <span property="value">doQuery</span>

</span>

<span property="generic-parameter" typeof="keyval">

   <span property="key">LANG</span>

   <span property="value">ADQL</span>

</span>

<span property="generic-parameter" typeof="keyval">

   <span property="key">QUERY</span>

   <span property="value">SELECT * from tap_schema.tables</span>

</span>
```

### 2.3.4  continuation property

If the examples are spread over multiple linked documents, the links to documents with additional examples **must** be within the parent element defining the *vocab* attribute and the link elements **must** contain the following additional attributes: a *property* attribute with the value continuation, a *resource* attribute with an empty value (referring to the current document), and the href attribute with the URL of another document formatted as above (i.e. another collection of examples that clients should read to collect the full set of examples).

```
<div vocab="ivo://ivoa.net/std/DALI#examples">

  <div id="x" resource="#x" typeof="example">

   <h2 property="name">Synchronous TAP query</h2>

     <p property="capability">ivo://ivoa.net/std/TAP/v1.0</p>

     <p property="generic-parameter" typeof="keyval">

       <span property="key">REQUEST</span>=<span

           property="value">doQuery</span>

     </p>

     <p property="generic-parameter" typeof="keyval">

       <span property="key">LANG</span>=<span

           property="value">ADQL</span>

     </p>

     <p property="generic-parameter" typeof="keyval">
```

```
      <span property="key">QUERY</span>=<span
        property="value">SELECT * from tap_schema.tables</span>
   </p>
</div>

<a property="continuation"
   href="simple_examples.html">Simple examples</a>
<a property="continuation"
   href="fancy_examples.html">Fancy examples</a>
</div>
```

In the above example, the two linked documents would also contain some element with a *vocab* and examples as described above.

## 2.4 Availability: VOSI-availability

VOSI-availability [9] defines a simple web resource that reports on the current ability of the service to perform. In DAL services, this resource is always accessed as a resource named *availability* that is a child of the base URL for the service.

All DAL services **must** implement the */availability* resource, which responds with a dynamically generated document describing the current state of the service operation, e.g.:

```
<?xml version="1.0" encoding="UTF-8"?>
<vosi:availability
  xmlns:vosi="http://www.ivoa.net/xml/VOSIAvailability/v1.0">
  <vosi:available>true</vosi:available>
  <vosi:note>service is accepting queries</vosi:note>
</vosi:availability>
```

## 2.5 Capabilities: VOSI-capabilities

VOSI-capabilities [9] defines a simple web resource that returns an XML document describing the service. In  DAL services, this resource is always accessed as a resource named *capabilities* that is a child of the base URL for the service. The VOSI-capabilities should describe all the resources exposed by the service, including which standards each resource implements.

All DAL services **must** implement the /*capabilities* resource. The following capabilities document shows the three VOSI resources and a TAP base resource:

```xml
<?xml version="1.0" encoding="UTF-8"?>
<vosi:capabilities
   xmlns:vosi="http://www.ivoa.net/xml/VOSICapabilities/v1.0"
   xmlns:xsi="http://www.w3.org/2001/XMLSchema-instance"
   xmlns:vod="http://www.ivoa.net/xml/VODataService/v1.1">
  <capability standardID="ivo://ivoa.net/std/VOSI#capabilities">
    <interface xsi:type="vod:ParamHTTP" version="1.0">
      <accessURL use="full">
         http://example.com/tap/capabilities
      </accessURL>
    </interface>
  </capability>
  <capability standardID="ivo://ivoa.net/std/VOSI#availability">
    <interface xsi:type="vod:ParamHTTP" version="1.0">
      <accessURL use="full">
         http://example.com/tap/availability
      </accessURL>
    </interface>
  </capability>
  <capability standardID="ivo://ivoa.net/std/VOSI#tables">
    <interface xsi:type="vod:ParamHTTP" version="1.0">
      <accessURL use="full">
        http://example.com/tap/tables
      </accessURL>
    </interface>
  </capability>
  <capability xmlns:tr="http://www.ivoa.net/xml/TAPRegExt/v1.0"
     standardID="ivo://ivoa.net/std/TAP"
xsi:type="tr:TableAccess">
    <interface xsi:type="vod:ParamHTTP" role="std" version="1.0">
      <accessURL use="full">
            http://example.com/tap/
      </accessURL>
    </interface>
```

```
    <!-- service details from TAPRegExt go here -->

  </capability>
</vosi:capabilities>
```

Note that while this example shows the use of a registry extension schema (the inline xmlns:tr="http://www.ivoa.net/xml/TAPRegExt/v1.0" in the last capability element) this is not required; services may be registered and described without such an extension. The use of standardID -- which should contain the IVORN of the standard a capability adheres to -- does not imply a particular (or any) xsi:type be included.

## 2.6 Content: VOSI-tables

VOSI-tables [9] defines a simple web resource that returns an XML document describing the content of the service. In DAL services which include it, this resource is always accessed as a resource named *tables* that is a child of the base URL for the service. The document format is defined by the VODataService-1.1 [12] standard and allows the service to describe their content as a tableset: schemas, tables, and columns. If the VOSI-tables resource is not implemented by a service, the service **must** respond to a GET request with an HTTP 404 (Not Found) status.

A concrete DAL service specification will specify if the */tables* resource is permitted or required. The current VOSI-tables specification has some scalablity issues for services with many or large tables, so that specification is subject to change in future. Since DAL services with a VOSI-tables resource will specify in the capabilities which version they are using, DAL services can make use of new versions without change to the DAL service specification.

# 3 Parameters

A DAL job is defined by a set of parameter-value pairs. Some of these parameters have a standard meaning and are defined here, but most are defined by the service specification or another related standard.

## 3.1 General Rules

### 3.1.1 Case Sensitivity

Parameter names are not case sensitive; a DAL service **must** treat upper-, lower-, and mixed-case parameter names as equal. Parameter values are case sensitive unless a concrete DAL service specification explicitly states that the values of a specific parameter are to be treated as case-insensitive. For example, the following are equivalent:

```
FOO=bar
Foo=bar
foo=bar
```

Unless explicitly stated by the service specification, these are not equivalent:

```
FOO=bar
FOO=Bar
FOO=BAR
```

In this document, parameter names are typically shown in upper-case for typographical clarity, not as a requirement.

### 3.1.2 Literal Values: Numbers, Boolean, Date, and Time

Integer numbers **must** be represented in a manner consistent with the specification for integers in *XML Schema Datatypes* [4].

Real numbers **must** be represented in a manner consistent with the specification for double-precision numbers in *XML Schema Datatypes [4]*. This representation allows for integer, decimal and exponential notations.

Boolean values **must** be represented in a manner consistent with the specification for Boolean in XML Schema Datatypes [10]. The values *0* and *false* are equivalent. The values *1* and *true* are equivalent.

```
FOO=1
FOO=true
```

```
BAR=0
BAR=false
```

Date and time values **must** be represented using the convention established for FITS [18] and STC [15]:

```
YYYY-MM-DD['T']hh:mm:ss[.SSS]
```

where the T is a character separating the date and time components. The time component is optional, in which case the T separator is not used. Fractions of a second are permitted but not required. For example:

```
2000-01-02T15:20:30.456

2001-02-03T04:05:06

2002-03-04
```

are all legal date or date plus time values. Values never include a time zone indicator and are always interpreted as follows. In cases where values may be expressed using Julian Date (JD) or Modified Julian Date (MJD), these follow the rules for double precision numbers above and may have additional metadata as described in [11]. All date-time values (formatted sting, JD, and MJD) shall be interpreted as referring to time scale UTC and time reference position UNKNOWN, unless either or both of these are explicitly specified to be different [15].

Note that the format used here is very close to the standard ISO8601 timestamp format except with respect to timezone handling. ISO8601 requires a Z character at the end of the string when the timezone is UTC; here, we follow the FITS [18] convention by omitting the Z but still defaulting to UTC.

### 3.1.3 Multiple Values

Parameters may be assigned multiple values with multiple parameter=value pairs using the same parameter name. Whether or not multiple values are permitted and the meaning of multiple values is specified for each parameter by the specification that defines the parameter. For example, the UPLOAD parameter (Section 3.2.5 ) permits multiple occurrences of the specified pair (table,uri), e.g.:

```
UPLOAD=foo,http://example.com/foo

UPLOAD=bar,http://example.com/bar
```

Services **must** respond with an error if the request includes multiple values for parameters defined to be single-valued.

## 3.2 Standard Parameters

### 3.2.1 REQUEST

The REQUEST parameter specifies the type of the DAL job at the highest level. In many cases, a service will have only one possible value. This parameter is still used in such cases so that future versions or non-standard (site-specific) features may support additional values.

A service **must** respond with an error if the REQUEST parameter is missing or the value is not recognised.

The REQUEST parameter is always single-valued.

### 3.2.2 VERSION

The VERSION parameter is used so the client can specify which version of the service standard they are using to make the request. This allows implementers to support multiple versions of a standard in a single web service and with a single resource for the DAL job list. If the client does not specify a value for the VERSION, the service **should** interpret the request using the rules and semantics of the latest version supported by the service. However, service implementers may decide to make an older version the default. If the client requires features or behaviour of a specific version, an explicit VERSION parameter should be used.

The value of the version advertised by the service and requested by the client follows the IVOA version number scheme [14]. Standards at the Working Draft (WD) or Proposed Recommendation (PR) stage include the publication date with the version number. For the purposes of this specification, these date tags are not supported, i.e. services only accept VERSION without dates. Clients can thus not request the behaviour of a specific WD or PR. This is intentional since implementation against unstable standards are not supposed to be stable.

A service **must** respond with an error if the caller requests an unsupported version using the VERSION parameter.

The VERSION parameter is always single-valued.

### 3.2.3 RESPONSEFORMAT

The RESPONSEFORMAT parameter is used so the client can specify the format of the response (e.g. the output of the job). For DALI-sync requests, this is the content-type of the response. For DALI-async requests, this is the content-type of the result resource(s) the client can retrieve from the UWS result list resource; if a DALI-async job creates multiple results, the RESPONSEFORMAT should control the primary result type, but details can be specific to individual service specifications. While the list of supported values are specific to a concrete service specification, the general usage is to support values that are content-types (mimetypes [5]) for known formats as well as shortcut symbolic values.

| table type | MIME type(s) | short form |
|---|---|---|
| VOTable | application/x-votable+xml | votable |

| | text/xml | |
|---|---|---|
| comma separated values | text/csv | csv |
| tab separated values | text/tab-separated-values | tsv |
| FITS file | application/fits | fits |
| pretty-printed text | text/plain | text |
| pretty-printed Web page | text/html | html |

In some cases, the specification for a specific format may be parameterised (e.g. the mimetype may include optional semi-colon and additional key-value parameters). A DAL service **must** accept a RESPONSEFORMAT parameter indicating a format that the service supports and **should** fail (Section 4.2 ) where the RESPONSEFORMAT parameter specifies a format not supported by the service implementation.

A concrete DAL service specification will specify any mandatory or optional formats as well as new formats not listed above; it may also place limitations on the structure for formats that are flexible. For example, a resource that responds with tabular output may impose a limitation that FITS files only contain FITS tables, possibly only of specific types (ascii or binary).

If a client requests a format by specifying the mimetype (as opposed to one of the short forms), the response that delivers that content **must** set that mimetype in the Content-Type header. This is only an issue when a format has multiple acceptable mimetypes (e.g. VOTable). This allows the client to control the Content-Type so that it can reliably cause specific applications to handle the response (e.g. a browser rendering a VOTable generally requires the text/xml mimetype). If the client requests a plain mimetype (e.g. not parameterised) and the mimetype does allow optional parameters, the service may respond with a parameterised mimetype to more clearly describe the output. For example, the text/csv mimetype allows two optional parameters: charset and header. If the request includes RESPONSEFORMAT=text/csv the response could have Content-Type text/csv or text/csv;header=absent at the discretion of the service. If the request specifies specific values for parameters, the response **must** be equivalent.

Individual DAL services (not just specifications) are free to support custom formats by accepting non-standard values for the RESPONSEFORMAT parameter.

The RESPONSEFORMAT parameter should not be confused with the FORMAT parameter used in many DAL services. The latter is generally used as a query parameter to search for data in the specified format; FORMAT and RESPONSEFORMAT have the same sense in TAP-1.0, but this is not generally the case.

The RESPONSEFORMAT parameter is always single-valued.

### 3.2.4 MAXREC

For resources performing discovery (querying for an arbitrary number of records), the resource **must** accept a *MAXREC* parameter specifying the maximum number of records to be returned. If *MAXREC* is not specified in a request, the service **may** apply a default value or **may** set no limit. The service may also enforce a limit on the value of MAXREC that is smaller than the value in the request. If the size of the result exceeds the resulting limit, the service **must** only return the requested number of rows. If the result set is truncated in this fashion, it **must** include an overflow indicator as specified in Section 4.4.1 .

The service **must** support the special value of *MAXREC=0.* This value indicates that, in the event of an otherwise valid request, a valid response be returned containing metadata, no results, and an overflow indicator (Section 4.4.1 ). The service is not required to execute the request and the overflow indicator does not necessarily mean that there is at least one record satisfying the query. The service **may** perform validation and may try to execute the request, in which case a MAXREC=0 request can fail.

The MAXREC parameter is always single-valued.

### 3.2.5 UPLOAD

The *UPLOAD* parameter is used to reference read-only external resources (typically files) via their URI, to be uploaded for use as input resources to the query. The value of the *UPLOAD* parameter is a resource name-URI pair. For example:

```
UPLOAD=table1,http://example.com/t1
```

would define an input named *table1* at the given URI. Resource names **must** be simple strings made up of alphabetic, numeric, and the underscore characters only and **must** start with an alphabetic character.

Services that implement *UPLOAD* **must** support *http* as a URI scheme (e.g. must support treating an *http* URI as a URL). A VOSpace URI (*vos:<something>*) is a more generic example of a URI that requires more service-side functionality; support for the *vos* scheme is optional.

To upload a resource inline, the caller specifies the UPLOAD parameter (as above) using a special URI scheme "param". This scheme indicates that the value after the colon will be the name of the inline content. The content type used is *multipart/form-data*, using a "file" type input element. The "name" attribute must match that used in the UPLOAD parameter.

For example, in the POST data we would have this parameter:

```
UPLOAD=table3,param:t3
```

and this content:

```
Content-Type: multipart/form-data; boundary=AaB03

[...]

--AaB03x

Content-disposition: form-data; name="t3"; filename="t3.xml"

Content-type: application/x-votable+xml

[...]

--AaB03x

[...]
```

If inline upload is used by a client, the client must POST both the UPLOAD parameter and the associated inline content in the same request. Services that implement upload of resources **must** support the *param* scheme for inline uploads.

In principle, any number of resources can be uploaded using the UPLOAD parameter and any combination of URI schemes supported by the service as long as they are assigned unique names in the request. For example:

```
UPLOAD=table1,http://example.com/t1.xml

UPLOAD=image1,vos://example.authority!tempSpace/foo.fits

UPLOAD=table3,param:t3
```

Services may limit the size and number of uploaded resources; if the service refuses to accept the upload, it **must** respond with an error as described in Section 4.2 . Specific service specifications specify how uploaded resources are referenced in other request parameters (for example, in a query), and interpreted.

### 3.2.6 RUNID

The service **should** implement the *RUNID* parameter, used to tag service requests with the identifier of a larger job of which the request may be part. The RUNID value is a string with a maximum length of 64 characters.

For example, if a cross match portal issues multiple requests to remote services to carry out a cross-match operation, all would receive the same *RUNID*, and the service logs could later be analysed to reconstruct the service operations initiated in response to the job. The service **should** ensure that *RUNID* is preserved in any service logs and **should** pass on the *RUNID* value in calls to other services made while processing the request.

The RUNID parameter is always single-valued.

# 4  Responses

All DAL service requests eventually result in one of three kinds of responses: successful HTTP status code (200) and a service- and resource-specific representation of the results, an HTTP status code and a standard error document (see below) or a  service- and resource-specific error document, or a redirect HTTP status code (302 or 303) with a URL in the HTTP Location header.

## 4.1  Successful Requests

Successfully executed requests **must** eventually (after zero or more redirects) result in a response with HTTP status code 200 (OK) and a response in the format requested by the client (Section  3.2.3 ) or in the default format for the service. The service should set the following HTTP headers to the correct values where possible.

| | |
|---|---|
| Content-Type | mimetype of the response |
| Content-Encoding | encoding/compression of the response |
| Content-Length | size of the response in bytes (generally not known for dynamically generated and streamed response) |
| Last-Modified | timestamp when the resource was last changed (not applicable to dynamically generated response) |

For jobs executed using a DALI-async resource, the result(s) **must** be made available as child resources of the result list and directly accessible there. For jobs that inherently create a fixed result, service specifications may specify the name of the result explicitly. For example, TAP-1.0 has a single result and it **must** be named *result* in the result list and be directly accessible by that name, e.g.:

```
GET http://example.com/base/joblist/123/results/result
```

For concrete DAL service specifications where multiple result files may be produced, the specification may dictate the names or it may leave it up to implementations to choose suitable names.

## 4.2  Errors

If the service detects an exceptional condition, it **must** return an error document with an appropriate HTTP-status code. DAL services distinguish three classes of errors:

• Errors in communicating with the DAL service

- Errors in the use of the specific DAL protocol, including an invalid request

- Errors caused by a failure of the service to complete a valid request

Error documents for communication errors, including those caused by accessing non-existent resources, authentication or authorization failures, services being off-line or broken are not specified here since responses to these errors may be generated by other off-the-shelf software and cannot be controlled by service implementations. There are several cases where a DAL service could return such an error. First, a DALI-async resource **must** return a 404 (not found) error if the client accesses a job within the UWS job list that does not exist, or accesses a child resource of the job that does not exist (e.g. the error resource of a job that has not run and failed, or a specific result resource in the result list that does not exist). Second, access to a resource could result in an HTTP 401 (not authorized) response if authentication is required or an HTTP 403 (forbidden) error if the client is not allowed to access the requested resource. Although UWS is currently specified for HTTP transport only, if it were to be extended for use via other transport protocols, the normal mechanisms of those protocols should be used.

An error document describing errors in use of the DAL service protocol **may** be a VOTable document as described in [11] or a plain text document. The content of VOTable error documents is described in Section 4.4.2 below. Service specifications will enumerate specific text to be included. For plain text error documents the required text would be included at the start of the document; for VOTable error documents, the required (and optional) text would be included as content of the INFO element described in Section 4.4.2 . In either case, DAL services will allow service implementers to add additional explanatory text after the required text (on the same line or on subsequent lines). In all cases, these are errors that occur when the job is executed and do not override any error behaviour for a UWS resource which specifies the behaviour and errors associated with interacting with the job itself.

If the invalid job is being executed using a DALI-async resource, the error document **must** be accessible from the <DALI-async>/<jobid>/error resource (specified by UWS) and when accessed via that resource it **must** be returned with an HTTP status code 200, e.g.:

```
GET http://example.com/base/joblist/123/error
```

For DALI-async errors, services should recommend and may mandate that required text be included in the error summary field of the UWS job in addition to the error document; this permits generic UWS clients to consume the standard part of the error description.

If the error document is being returned directly after a DALI-sync request, the service **should** use a suitable error code to describe the failure and include the error document in the body of the response. The Content-Type header will tell the client the format of the error document that is included in the body of the response. In general, HTTP status codes from 400-499 signify a problem with the client request and status codes greater than or equal to 500 signify that the request is (probably) valid but the server has failed to function. For transport protocols other than HTTP, the normal error reporting mechanisms of those protocols should be used.

## 4.3  Redirection

A concrete DAL service specification may require that HTTP redirects (302 or 303) be used to communicate the location of an alternate resource which should be accessed by the client via the HTTP GET method. For example, the UWS pattern used for DALI-async (Section  2.1 ) requires this behaviour. Even when not required, concrete DAL service specifications **must** allow implementers to use redirects and clients must be prepared to follow these redirects using normal HTTP semantics [5].

## 4.4  Use of VOTable

VOTable is a general format. In DAL services we require that it be used in a particular way. The result VOTable document **must** comply with VOTable v1.2 [11] or later versions.

The VOTable format permits table creators to add additional metadata to describe the values in the table. Once a standard for including such metadata is available, service implementers should use such mechanisms to augment the results with additional metadata. Concrete DAL service specifications may require additional metadata of this form.

The VOTable **must** contain one  *RESOURCE* element identified with the attribute *type="results"; this resource contains the primary result (e.g. the only result for simple DAL services)*. Concrete DAL service specifications define what goes into the primary result. The primary *RESOURCE* element **must** contain, before the *TABLE* element, an *INFO* element with attribute *name* = *"QUERY_STATUS"*[2]. The *value* attribute **must** contain one of the following values:

| OK | the job executed successfully and the result is included in the resource |
|----|-----|

---

[2]  QUERY_STATUS here simply means the status of the request and does not necessarily imply a query in the database sense. This name is used for backwards compatibility with earlier standards.

| ERROR | an error was detected at the level of the protocol, the job failed to execute, or an error occurred while writing the table data |
|-------|--------------------------------------------------------------------------------------------------------------------------------|
| OVERFLOW | the job executed successfully, the result is included in the resource, and the result was truncated at MAXREC rows |

The content of the INFO element conveying the status **should** be a message suitable for display to the user describing the status.

```
<INFO name="QUERY_STATUS" value="OK"/>

<INFO name="QUERY_STATUS" value="OK">Successful query</INFO>

<INFO name="QUERY_STATUS" value="ERROR">

value out of range in POS=45,91

</INFO>
```

Additional *RESOURCE* elements may be present, but the usage of any such elements is not defined here. Concrete DAL service specifications may define additional resources (and the type attribute to describe them) and service implementers are also free to add their own.

### 4.4.1 Overflow

If an overflow occurs (result exceeds MAXREC), the service **must** include an INFO element in the RESOURCE with *name="QUERY_STATUS"* and the value=*"OVERFLOW"*. If the initial info element (above) specified the overflow, no further elements are needed, e.g.:

```
<RESOURCE type="results">

<INFO name="QUERY_STATUS" value="OVERFLOW"/>

...

<TABLE>...</TABLE>

</RESOURCE>
```

If the initial info element specified a status of OK then the service **must** append an INFO element for the overflow after the table, e.g.:

```
<RESOURCE type="results">

<INFO name="QUERY_STATUS" value="OK"/>

...

<TABLE>...</TABLE>

<INFO name="QUERY_STATUS" value="OVERFLOW"/>

</RESOURCE>
```

In the above examples, the TABLE should have exactly MAXREC rows.

### 4.4.2 Errors

If an error occurs, the service **must** include an INFO element with name="*QUERY_STATUS*" and the value="*ERROR*". If the initial info element (above) specified the error, no further elements are needed, e.g.:

```
<RESOURCE type="results">
<INFO name="QUERY_STATUS" value="ERROR"/>
...
<TABLE>...</TABLE>
</RESOURCE>
```

If the initial info element specified a status of OK then the service **must** append an INFO element for the error after the table, e.g.:

```
<RESOURCE type="results">
<INFO name="QUERY_STATUS" value="OK"/>
...
<TABLE>...</TABLE>
<INFO name="QUERY_STATUS" value="ERROR">
unexpected IO error while converting something
</INFO>
</RESOURCE>
```

The use of trailing INFO element allows a service to stream output and still report overflows or errors to the client. The content of these trailing INFO elements is optional and intended for users; client software **should not** depend on it.

### 4.4.3 Additional Information

Additional *INFO* elements **may** be provided, e.g., to echo the input parameters back to the client in the query response (a useful feature for debugging or to self-document the query response), but clients **should not** depend on these. For example:

```
<RESOURCE type="results">
...
<INFO name="standardID" value="ivo://ivoa.net/TAP"/>
<INFO name="standardVersion" value="1.0"/>
...
</RESOURCE>
```

The following names for INFO elements should be used if applicable, but this list is not definitive.

| name | meaning |
|------|---------|
| standardID | IVOA standardID for the service specification |
| standardVersion | Version number used to interpret the request (Section  3.2.2 ) |
| citation | Reference to a publication that can/should be referenced if the result is used |

For citations, the INFO element should also include a *ucd* attribute with the value *meta.bib* (if the value is a free-text reference) or *meta.bib.bibcode* (if the value is a bibcode). If other meta.bib UCDs are added to the vocabulary in future, they may also be used to describe the value.

# 5 Changes

## 5.1 PR-DALI-20130919

The following changes are in response to additional RFC commands and during the TCG review.

New architecture diagram and minor editorial changes to improve document.

Clarified RESPONSEFORMAT text to allow services to append mimetype parameters if the client did not specify them.

Relaxed the VERSION parameter so services should default to latest (instead of must) and to not differentiate between REC and pre-REC status.

Clarified the requirement for a VOTable RESOURCE with type="results" attribute so it is clear that this is the primary result and other RESOURCES may be present.

Clarified that HTTP-specific rules apply to RESTful web services and that although we describe such services here we do not preclude future use of other transport protocols.

## 5.2 PR-DALI-20130521

The following changes are in response to comments from the RFC period.

Made editorial changes from the DALI wiki page that were missed during WG review.

Changed all cross-references to be readable text.

Replaced example curl output from a POST with explanatory text.

POST of job parameters directly to job: restricted to creation and /parameters resource

Changed number of DALI-async and DALI-sync resources to zero or more.

Clarified that job parameters are the same if the type of job is the same, but services can have different types of jobs (and hence different parameters) on different job-list resources.

Fixed text forbidding any other vocab attributes in DALI-examples document.

Replace http-action and base-url with something or add sync vs async: replaced with capability property

Preventing loops with continuation in examples: removed.

Clarified that VOSI-capabilities does not require a registry extension schema and use of xsi:type.

Explicitly require that if VOSI-tables is not implemented, the service responds with a 404.

Clarified the purpose of requiring the service to use client-specified RESPONSEFORMAT as the Content-Type of the response.

Attempted to clarify the acceptable use of status codes for errors.

Removed single-table restriction from votable usage.

Clarified interpretation of dates and times as UTC timescale by default but permitting specific metadata to be specified.

Removed formatting of example links so they are not real hyperlinks in output documents.

Clarified that services can enforce a smaller limit than a requested MAXREC.

Removed text refering to IVOA notes on STC and Photometric metadata; added more general text that services should include additional metadata once standards for such are in place.

Explain the table at start of section 2.

Clarify requests that effect UWS job phase in DALI-async.

Removed malformed http post example from DALI-async section.

Remove reference to SGML specifically, but mention HTML5 as a poor choice for DALI-examples.

Add reference to RFC2616 in the RESPONSEFORMAT section since it talks about mimetypes.

Clarified text about setting job parameters and banned posting parameters directly to the job URL.

Replaced the base-url and http-action properties with a single capability property in DAL-examples. Changed the vocab identifier to be the ivorn for DALI with fragment indicating the DALI-examples section of the document.

### 5.3  WD-DALI-1.0-20130212

Simplified DALI-examples to conform to RDFa-1.1 Lite in usage of attributes.